\documentclass[twoside,twocolumn,english,aps,showpacs,prl,superscriptaddress]{revtex4-1}
\usepackage[T1]{fontenc}
\usepackage[latin9]{inputenc}
\usepackage{color}
\usepackage{bm}
\usepackage{amsmath}
\usepackage{amssymb}
\usepackage{graphicx}
\usepackage{esint}
\usepackage{mathrsfs}
\usepackage{float}
\usepackage{wasysym}
\usepackage{amssymb}
	\usepackage{pdfpages}
	\makeatletter
	\AtBeginDocument{\let\LS@rot\@undefined}
	\makeatother

\makeatletter

\makeatletter
\newcommand*{\rom}[1]{\expandafter\@slowromancap\romannumeral #1@}
\makeatother

\usepackage{times}{\Large }

\makeatother

\usepackage{babel}
\begin{document}
\title{A hierarchy in Majorana non-abelian tests and hidden variable models}

\author{Peng Qian}
\affiliation{State Key Laboratory of Low Dimensional Quantum Physics, Department of Physics, Tsinghua University, Beijing, 100084, China}
\affiliation{Beijing Academy of Quantum Information Sciences, Beijing 100193, China}

\author{Dong E. Liu}
\email{Corresponding to: dongeliu@mail.tsinghua.edu.cn}
\affiliation{State Key Laboratory of Low Dimensional Quantum Physics, Department of Physics, Tsinghua University, Beijing, 100084, China}
\affiliation{Beijing Academy of Quantum Information Sciences, Beijing 100193, China}
\affiliation{Frontier Science Center for Quantum Information, Beijing 100184, China}


\begin{abstract}
The recent progress of the Majorana experiments paves a way for the future tests of non-abelian braiding statistics and topologically-protected quantum information processing. However, a deficient design in those tests could be very dangerous and reach false-positive conclusions. A careful theoretical analysis is necessary in order to develop loophole-free tests. We introduce a series of classical hidden variable models to capture certain key properties of Majorana system: non-locality, topologically non-triviality, and quantum interference. Those models could help us to classify the Majorana properties and to set up the boundaries and limitations of Majorana non-abelian tests: fusion tests, braiding tests and test set with joint measurements. We find a hierarchy among those Majorana tests with increasing experimental complexity.
\end{abstract}
\maketitle

{\em Introduction--}The building blocks of topological quantum computation \cite{toricKitaev,freedman2003topological,NayakRMP} are non-abelian anyons, which are proposed to show novel non-Abelian braiding statistics \cite{Leinaas&Myrheim77,fredenhagen1989,Ivanov,NayakRMP}. A simple type of non-abelian anyon, i.e. Majorana zero mode (MZM), was theoretically proposed~\cite{ReadGreen,1DwiresKitaev,Fu&Kane08,SatoPRL09,Sau10,LutchynPRL10,1DwiresOreg,Alicea10,AliceaRev} in many physically realizable systems. Recent experimental progresses makes an almost conclusive observation of Majorana resonance\cite{Mourik2012,Deng2012,Churchill2013,Finck2013,Nadj-Perge14,Albrecht16,deng2016Majorana,Zhang2017Ballistic,Sun2016Majorana,Wang2018Evidence,TamegaiVortexMZM2019,Liu2018Robust,Fornieri2019NatureTJJ,Ren2019Nature}. Even though it is necessary to build up Majorana devices step-by-step~\cite{ZhangLiuReview} experimentally, these progresses makes a promising platform towards the test of non-abelian braiding statistics and the actual quantum information processing.
The most popular schemes proposed for validating non-abelian statistics are braiding tests and fusion tests~\cite{NayakRMP,aasen2016milestones,ClarkeBraiding,van2012coulomb}. Besides, joint-measurement, C-NOT gate, non-trivial T-gate test and Bell non-locality test ~\cite{Bell1966,chsh1969} are believed to be necessary for quantum information processing. 
Here, we focus on relatively simple properties: Majorana non-local behaviors, topologically non-triviality, and quantum interference and entanglement in the near future. 

In the hidden variable theories, the probabilities appeared when we measure physical quantities are due to the lack of the knowledge or statistical approximation in the underlying complex deterministic theory\cite{Bohm52,Everett53,AspectPRL82,GISIN1991201,Mermin93,Bassi03,Scott03,GENOVESE2005,leggett2006nonlocal,Spekkens2007,Augusiak_2014}, which reproduces certain behaviors in quantum mechanics. The experimental activities in the Bell inequality test eventually rule out local hidden variable explanations in several physical systems \cite{aspect1982experimental,weihs1998violation,simon2003robust,garcia2004proposal,colbeck2011no,dada2011experimental,pusey2012reality,hensen2015loophole,hensen2016loophole,rosenfeld2017event}. But, this is not sufficient to say that the behaviors observed in other systems, e.g. Majorana platforms, are quantum mechanical.
For our purpose, if we want to validate Majorana behaviors seriously for future quantum information applications, it is reasonable to treat the system as a black-box and perform the test subsequently. In that sense, potential hidden variable theories need to be carefully considered and excluded. Otherwise, a deficient test design might reach a positive but incorrect conclusion. For example, braiding and measurements in Majorana systems can generate topologically protected Clifford gates including entanglement gates, which can also be realized in certain hidden variable theories~\cite{Spekkens2007}. Therefore, it is natural to ask whether we can find certain hidden variable theories such that each theory can capture certain properties of Majorana systems but can not capture other tests' results; and the hope is to set up the limitations and boundaries of each Majorana test by using these theories.

In this work, we proposed a couple of  classical hidden variable (HV) models, which are designed to capture and/or distinguish certain key quantum mechanical properties of Majorana systems: simplest Majorana non-local behaviors, topologically non-triviality and quantum interference. We want to show what properties can be seen in a particular test; and examine if the test results can be captured by both theories, which help us to set up the boundaries and limitations of the corresponding test. Based on this philosophy, we show:1) fusion tests only capture certain Majorana non-local behaviors but cannot capture others; 2) braiding tests show non-abelian statistics indicating topologically non-triviality, but not necessarily quantum interference; 3) test set with joint measurements are helpful and necessary to capture quantum interference in Majorana systems. Finally, we conclude that there exists a hierarchy among the Majorana non-abelian tests: Fusion test, braiding test, and test set with joint measurements.

{\em Theoretical Framework--}
Now, we introduce a couple of HV models which are inspired from Spekkens' famous model~\cite{Spekkens2007}.

\textbf{Classical HV theory \uppercase\expandafter{\romannumeral1}:} We assume that (1) we have incomplete knowledge about the system states for some unknown complexity, (2) we get an outcome and observe incomplete knowledge after specific operations called measurements, (3) we cannot get access to their hidden complexity. 

\begin{figure}[t]
\centerline{\includegraphics[scale=0.4]{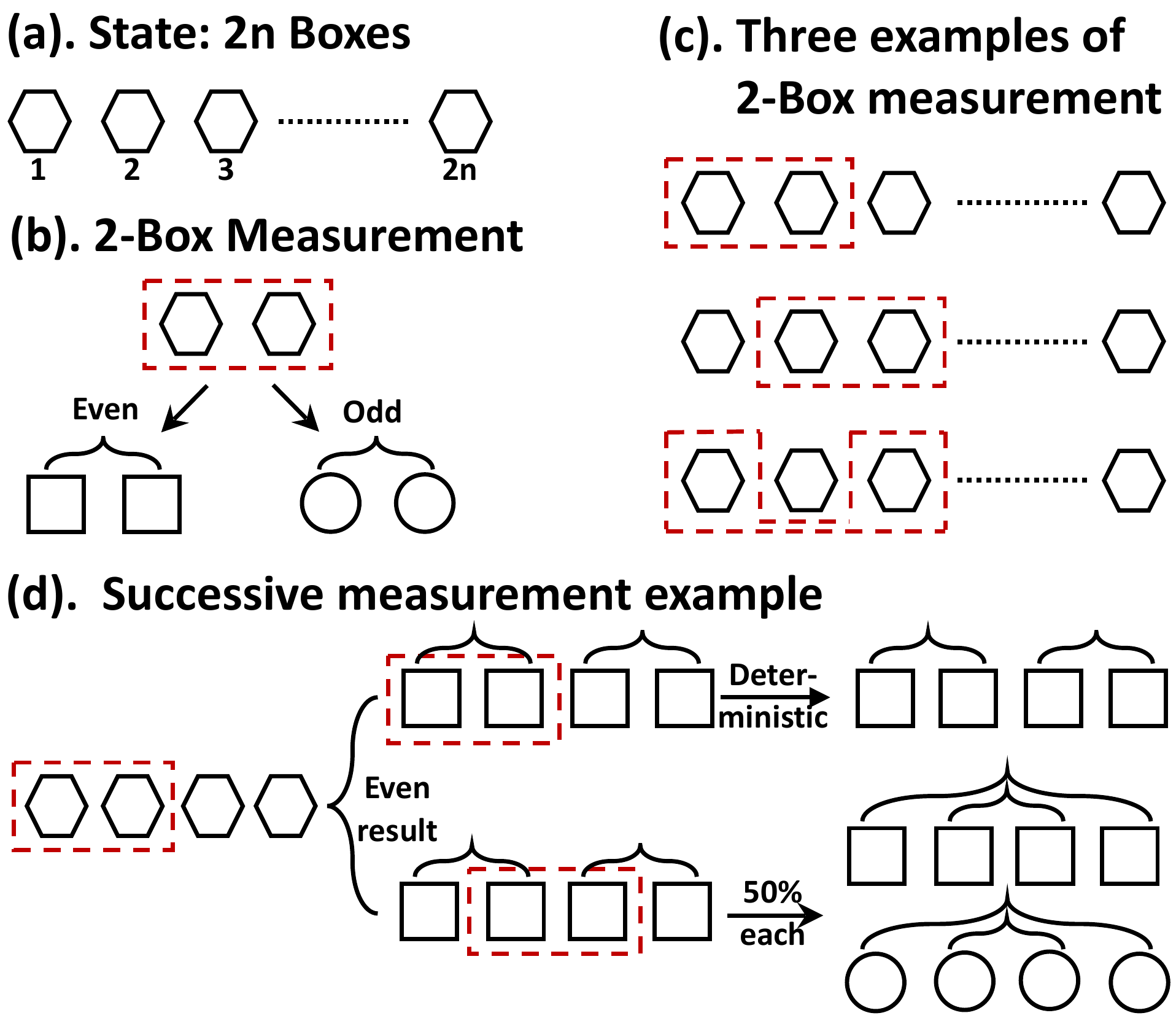}}
\caption{(a-c)The HV model \uppercase\expandafter{\romannumeral1}, where the number $1$ to $2n$ labels the position of boxes from left to right. (d) A four-box system (with total parity even) for fusion. If we get even parity result after Box 1,2-measurement, then continuing to apply Box 1,2-measurement, the state is unchanged.  But if we apply a Box 2,3-measurement in the second step, we have two possible outcomes with equal probability.}
\label{fig:MajoranaBox}
\end{figure}

a). State formalism--We consider a classical theory with a box representation, where each box corresponds to a MZM as shown in Fig.\ref{fig:MajoranaBox}; and the box can be either filled or empty. We assume that we can only know the parity of any (pair of) two boxes after measurement. For example, we measure the combination of the first and the second boxes and then get the result even. We can only know that the two boxes are either both filled or empty, but cannot determine the exact state of each box. This uncertainty is due to the unobserved classical complexity corresponding to the hidden-variables. 

b). Measurement--We show three kinds of 2-box measurements (Box 1,2, Box 2,3, and Box 1,3) in Fig.\ref{fig:MajoranaBox}:. Each measurement returns two-box parity: even or odd. We also assume the parity of the all-box system is fixed similar to the Majorana systems. After Box 1,2-measurement for example, we could either obtain an even-parity result $\overbrace{\square\square}$ with a pair of square-symbols or obtain an odd parity result $\overbrace{\Circle\Circle}$ with a pair of circle-symbols; we use a bracket to label post-measurement box-pair called "connections". Considering a four-box (with total parity even) example shown in Fig.\ref{fig:MajoranaBox} (d), Box1,2-measurement will reach either an even-even state: $\overbrace{\square\square}\overbrace{\square\square}$, or an odd-odd state:$\overbrace{\Circle\Circle}\overbrace{\Circle\Circle}$ with equal probability. If we repeat the previous measurement, we will get the same result;  if the measurement is different from the previous one, we will get an uncertain result. 

\begin{figure}[t]
\centerline{\includegraphics[scale=0.45]{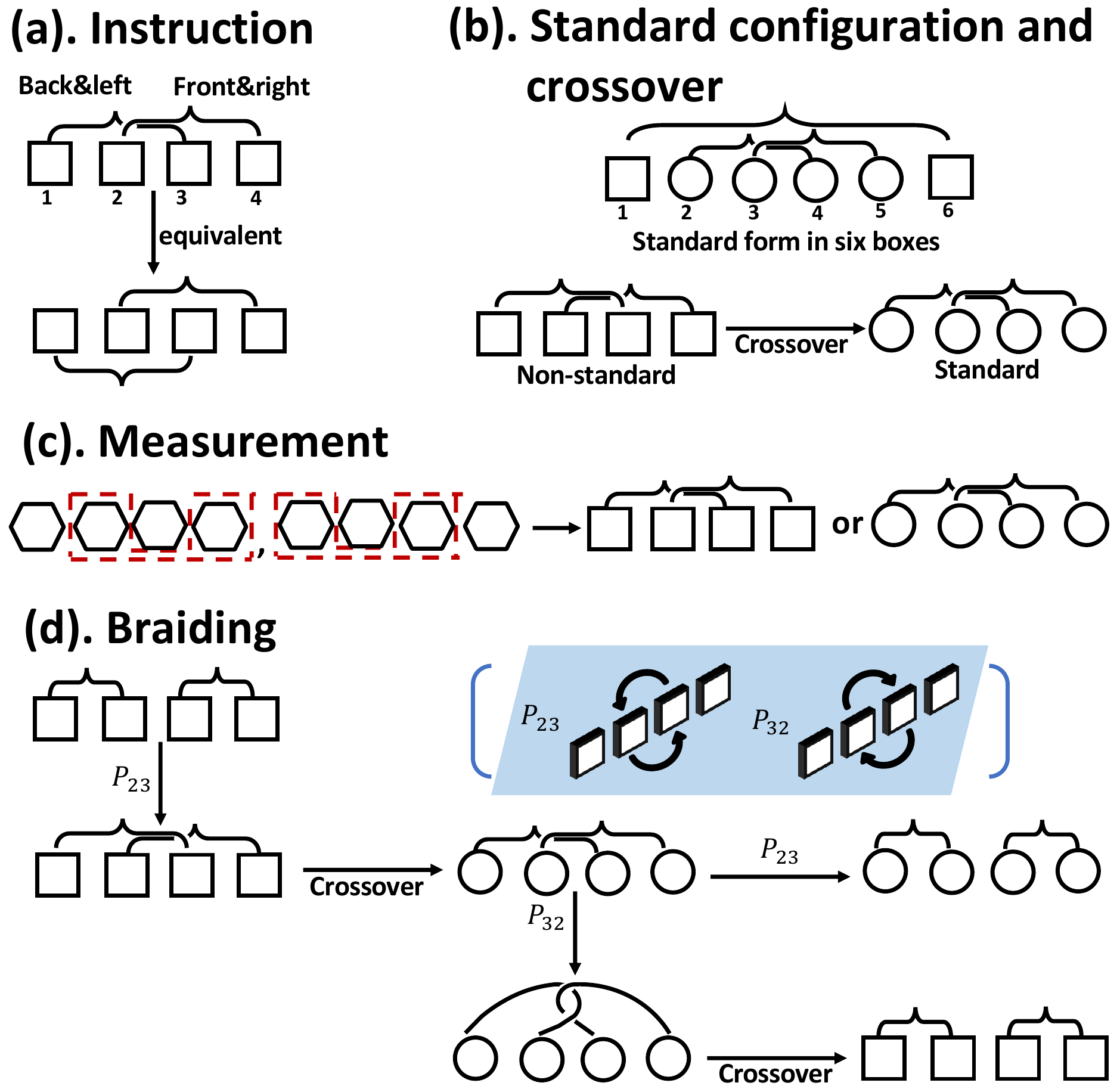}}
\caption{The HV model \uppercase\expandafter{\romannumeral2}. (a) Two equivalent form of state description: the lower means the back. (b) The standard configuration of a six-box state: the connection of Box1,6 is both the rightmost and frontmost as definition, and connection of Box3,5 is relative right and front to Box2,4. For non-standard configurations, we can apply crossover along with a parity change. (c) Both Box1,3 and Box2,4 measurements lead to standard configurations. (d) Braiding procedures $P^2_{23}$ and $P_{23}P_{32}$ . The braiding is confined in the plane perpendicular to the paper, or in the light-blue plane shown in the bracket.}
\label{fig:MajoranaBox2}
\end{figure}

\textbf{Classical HV theory \uppercase\expandafter{\romannumeral2}:}  To include topology, we take into account the relative positions of connections if two connections have overlaps as shown in Fig.\ref{fig:MajoranaBox2} (a): (1) the front-back order (front connections are on the top of back ones) of the connections has a non-trivial consequence; (2) we also define the right-left order of connections, where the connection is the right as long as one box in the pair is in the right of all boxes in other box-pairs. 

a). Standard states--For the order of connections discussed above, we can choose any order. However, for the convenience of discussion, we define a standard configuration with fixed order: the right connection is the front connection (e.g. in the upper part of Fig.\ref{fig:MajoranaBox2} (b)). For non-standard cases, we need to apply crossovers (exchange positions of two overlapping connections) to reach a standard configuration. Each crossover will induce a parity change (an example shown in Fig.\ref{fig:MajoranaBox2} (b)). In the text, we sometimes use an up-down bracket to describe an equivalent front-back position, e.g. in Fig.\ref{fig:MajoranaBox2} (a).

b). Measurement--We only consider measurements in standard configurations, i.e. "standard measurement rule". For an unidentified state, measurements (e.g. both Box2,4 and Box1,3 measurements in Fig.\ref{fig:MajoranaBox2} (c)) always result in standard configurations. On the other hand, if we know the connections of boxes, we only apply the standard rule for the standard cases. If we're in a non-standard form, we need to apply crossovers to reach a standard one.

c). Braiding--In our theory, braiding corresponds to the position exchange of boxes in the plane shown in Fig.\ref{fig:MajoranaBox2} (d). Considering different directions of braiding, the exchange of Box $n_1$ and Box $n_2$ is labeled by $P_{n_1n_2}$ for anticlockwise exchange if $n_1<n_2$ (for clockwise exchange if $n_1>n_2$). In an example in Fig.\ref{fig:MajoranaBox2} (d), we apply two paths of braiding operations on the same initial state $\overbrace{\square\square}\overbrace{\square\square}$. The first path contains two successive $	P_{23}$ where the first $P_{23}$ takes the state to a non-standard form and we apply a crossover to reach the standard one, then applying the second $P_{23}$ to get $\overbrace{\Circle\Circle}\overbrace{\Circle\Circle}$. In the second path, the first part is the same while we apply $P_{32}$ in the next step and we get a non-standard form with a knot. To reach a standard configuration, we apply a crossover to remove the knot and obtain the form $\overbrace{\square\square}\overbrace{\square\square}$. \\

{\em Majorana non-Abelian Tests--}Let us now consider a variety of Majorana tests, and check if their key properties, e.g. Majorana non-local behaviors, topologically non-triviality, and quantum interference, can be captured by our classical HV theory \uppercase\expandafter{\romannumeral1} and \uppercase\expandafter{\romannumeral2}.

A). \textbf{Fusion test:} Non-abelian anyon system follow special fusion processes which describe the outcomes after anyon combinations \cite{NayakRMP}. Majorana zero modes belong to the Ising non-abelian anyon model, which includes three types of anyons: the vacuum $I$, non-abelian anyon $\sigma$ (i.e. Majorana), and the fermion $\psi$. A pair of $\sigma$ combine to fuse into either a vacuum or a fermion: $\sigma\times\sigma\rightarrow I+ \psi$. With more anyons, we have multiple ways to fuse them together, where the quantum states describing fusion transformation can be written as:
\begin{equation}
|a,b\rightarrow i\rangle|i,c\rightarrow d\rangle=\sum_{j}(F^{d}_{abc})^{i}_{j}|b,c\rightarrow j\rangle|a,j\rightarrow d\rangle,
\end{equation}
where $|a,b\rightarrow i\rangle|i,c\rightarrow d\rangle$ indicates a state with particular fusion channel where anyons $a$ and $b$ first fuse to $i$, and then $i$ and $c$ fuse to $d$. The matrix $(F^{d}_{abc})^{i}_{j}$ describes the transformation under different fusion channels, For Ising anyons, $F^{\sigma}_{\sigma\sigma\sigma}=\frac{1}{\sqrt 2}\left(
\begin{array}{cc}
1 & 1 \\
1 & -1 \\
\end{array}
\right).
$

\begin{figure}[t]
\centerline{\includegraphics[scale=0.53]{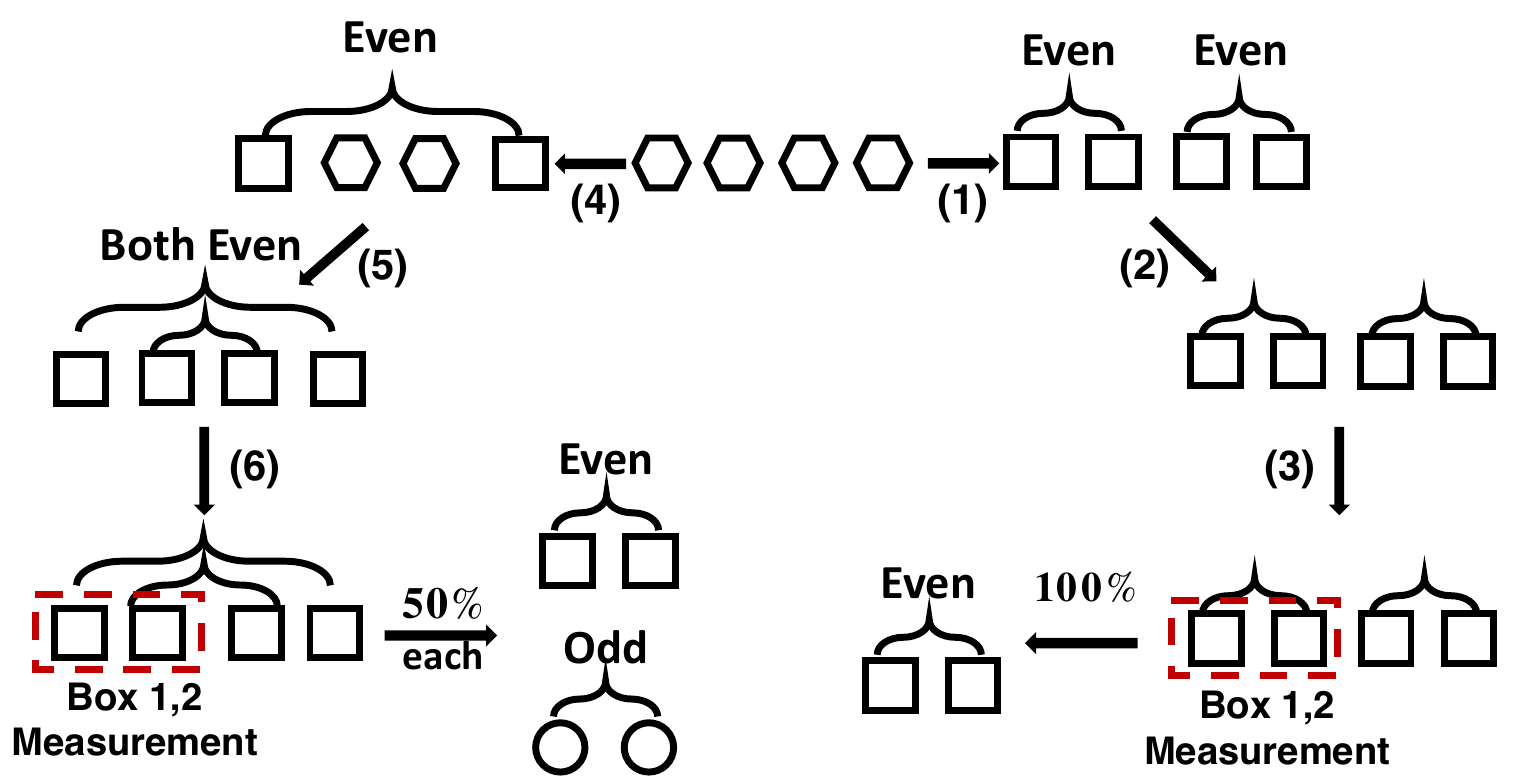}}
\caption{Fusion process of our HV model reproducing the Majorana fusion processes in~\cite{aasen2016milestones}. In the right path steps (1-3): the 4-box state splits into two pairs , both 2-box measurements obtain even, and then the measurement in the same box-pair Box1,2 obtains deterministic even parity result.  In the left path steps (4-6): the 4-box state splits into two different pairs (Box1,4 and Box2,3) with even parity after measurement, and then the  Box1,2  measurement gives us an equal probability of even or odd result.}
\label{fig4}
\end{figure}

Now, let's first focus on our HV theory \uppercase\expandafter{\romannumeral1}, we have:
\begin{eqnarray}
\hexagon\hexagon\hexagon\hexagon\rightarrow\overbrace{\square\square}\overbrace{\square\square}\rightarrow(\overbrace{\square\overbrace{\square\square}\square}+\overbrace{\Circle\overbrace{\Circle\Circle}\Circle})/2 \label{eq:HV-fusion1}\\
\hexagon\hexagon\hexagon\hexagon\rightarrow\overbrace{\Circle\Circle}\overbrace{\Circle\Circle}\rightarrow(\overbrace{\square\overbrace{\square\square}\square}+\overbrace{\Circle\overbrace{\Circle\Circle}\Circle})/2\label{eq:HV-fusion2}
\end{eqnarray}
Here, the first arrow in both Eq. (\ref{eq:HV-fusion1}) and (\ref{eq:HV-fusion2}) indicates the fusion (measurement) of Box$1,2$, which yields either fixed even parity (vacuum $I$) or odd parity (fermion $\psi$); the following Box2,3 measurement (second arrow) can have two possibilities: Box$2$ and Box$3$ first fuse into even parity or odd parity. We see that the measurements in HV model are analogous to the fusion, and reach the same measurement statistics as those in Ising anyon model described by the F-matrix above. Considering a proposed fusion test in~\cite{aasen2016milestones}, the topological superconducting device can realize two different 4-MZM fusion routes: 1) first create two pairs of MZMs and then fuse in the same pair as their creation, 2) first create two pairs of MZMs and then fuse in the different pairs. These two fusion routes gives different measurement statistics. As shown in Fig.\ref{fig4}, it is simple to check that the corresponding fusion procedures in our HV model generate the same measurement outcome statistics as those in Majorana cases. Our HV model can fully capture the measurement outcome after fusion-only processes in Majorana systems (see part \uppercase\expandafter{\romannumeral1} of SI \cite{supp} for more examples). 
This HV theory \uppercase\expandafter{\romannumeral1} does not include any topological consideration. Therefore, there is no doubt that the fusion tests can only capture certain Majorana non-locality behaviors but not including topological non-triviality.

B.) \textbf{Braiding and topology:} Non-Abelian anyons show nontrivial behaviors under the braiding process, and the exchange operation of anyons $a$ and $b$ can be captured by the operator $B_{ab}$, where these braiding operations can be written as $B_{ij}\propto1+\gamma_{i}\gamma_{j}$ for Majorana modes \cite{NayakRMP}.

Let us consider whether our HV theory \uppercase\expandafter{\romannumeral1} can capture braiding relations. For an example, braiding Box2,3 of initial state  $\overbrace{\square\square}\overbrace{\square\square}$ requires the exchange of boxes in different box-pairs, as in different fusion channel of Majorana theory. After this braiding, we get $\rlap{$\overbrace{\phantom{\square\square\square}}$}\square\overbrace{\square\square\square}$ (no connection order is considered here), then the  Box1,3 measurement (correspondence to $\sigma_{y}$ measurement in quantum case \cite{supp}) results in even parity. However, the $\sigma_{y}$ measurement of quantum state $B_{23}|0\rangle$ reaches odd. Therefor we fail to capture the braiding process without topological consideration. Now let's look at HV theory \uppercase\expandafter{\romannumeral2}. Considering the the same process on the initial state $\overbrace{\square\square}\overbrace{\square\square}$, we reach the state $\overbrace{\Circle\Circle}\overbrace{\Circle\Circle}$ as shown in the first path in Fig.\ref{fig:MajoranaBox2} (d), which is the same as quantum case.

\begin{figure}[t]
\centerline{\includegraphics[scale=0.5]{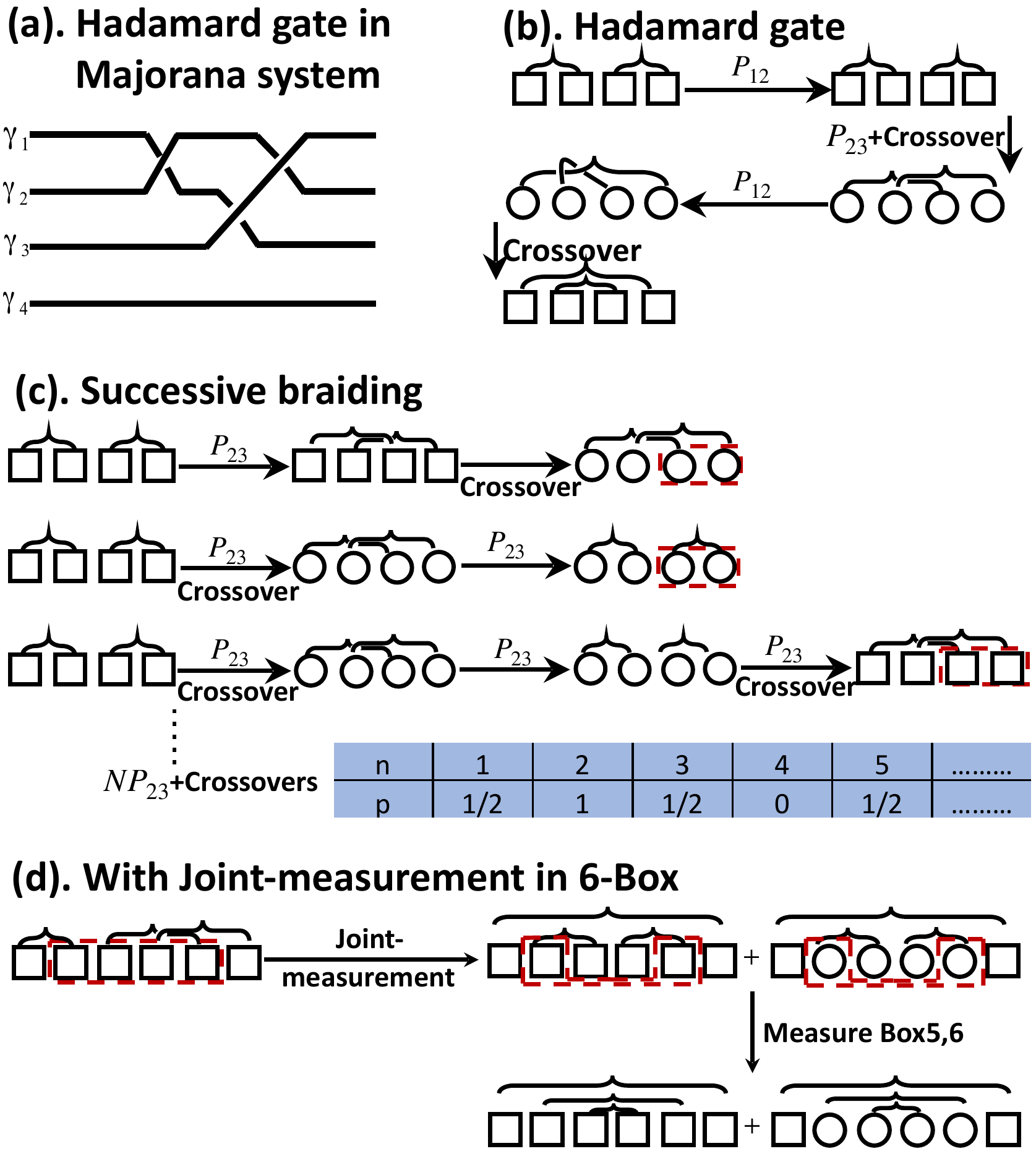}}
\vspace{-8pt}
\caption{(a). Hadamard gate in majorana-based quantum computation. (b). Hadamard gate in our HV theory \uppercase\expandafter{\romannumeral2} wherein the second step and last step we have applied crossovers to change into a standard configuration. (c). Example of non-abelian braiding in four boxes case in our HV theory \uppercase\expandafter{\romannumeral2}. Suppose in the initial state Box3,4 are in even parity. The table shows the possibility $p$ to obtain the odd parity in Box3,4 measurement, where the label $n$ describes the number of successive braiding procedures of Box2 and Box3.}\label{fig:Topology}
\end{figure}

We further examine the Hadamard gates which can be realized using a sequence of Majorana braidings $H=B_{12}B_{23}B_{12}$ in the quantum anyon model as shown in Fig.\ref{fig:Topology} (a). For our HV theory \uppercase\expandafter{\romannumeral2} in Fig.\ref{fig:Topology} (b), we also successively braid Box1,2($P_{12}$), Box2,3($P_{23}$) and Box1,2($P_{12}$) and then apply a crossover to get a standard form, and then we can simply show any measurement here gives the same results as those in Majorana Hadamard gate. For the complete cases please refer to part \uppercase\expandafter{\romannumeral2} of SI \cite{supp}.

Let's consider the "successive braiding", i.e. a non-abelian braiding test \cite{hyart13} with four Majorana modes $\gamma_1,\gamma_2,\gamma_3,\gamma_4$ where both $\gamma_1,\gamma_2$ and $\gamma_3,\gamma_4$ pair are initialized in even parity. We apply multiple braiding procedures ($n$ times) of $\gamma_2$ and $\gamma_3$ in an anti-clockwise way. We focus on the probability to obtain the odd result in the $(\gamma_3 \gamma_4)$ parity measurement at different $n$s. Braiding once, the parity of $\gamma_3,\gamma_4$ has $50\%$ chance to be odd in measurement. After two successive braidings, we get the definite odd parity; after three successive braidings, we will get an equal probability of both results; after four successive braidings, we get the definite even parity. As the number $n$ increases, the probability will repeat this pattern with period 4. In our HV theory \uppercase\expandafter{\romannumeral2}, as shown in Fig.\ref{fig:Topology} (c), we get the same measurement consequence as those in quantum case.

In either Majorana model or our HV theory \uppercase\expandafter{\romannumeral2}, the objects (MZMs or boxes) describing a specific fusion channel are all paired. Therefore, the braiding of two objects involves at most two pairs that is four MZMs or boxes. Considering the braiding of two boxes among $2n$ boxes, this procedure only changes the topology of two connections (of two box-pairs) involved in the braiding, but can not change the relative positions between these two connections and the other $n$-2 connections. Therefore, any single braiding step in more box-cases involves only four relevant boxes. Examples for braiding in more-MZM systems can be found in part \uppercase\expandafter{\romannumeral2} of SI \cite{supp}. In summary, the braidings can be well described in our HV theory \uppercase\expandafter{\romannumeral2} with topology considerations; and this theory goes beyond the fusion tests and show the unique topological-nontrivial property of braiding operations.

C.) \textbf{Joint-measurement and quantum interference:} Let us assume there are two Majorana qubits, where each one is encoded into four MZMs; and the single-qubit Pauli measurements can be realized by fermion-parity measurement of different Majorana pairs within each qubit~\cite{NayakRMP}. The joint-measurement can be realized by the fermion-parity measurement of four MZMs, in which two MZMs are from one qubit and the other two are from the other qubit~\cite{NayakRMP}. For joint-measurement $ZZ$, the state $\frac{|0\rangle+|1\rangle}{\sqrt{2}}\otimes \frac{|0\rangle+|1\rangle}{\sqrt{2}}$ will collapses onto $(|00\rangle+|11\rangle)/\sqrt{2}$ if the measurement outcome is even, and thus generate entanglement between the two qubits. In our HV theory \uppercase\expandafter{\romannumeral2} with two "qubits" (eight boxes), we can measure the four-box joint-parity, in which two boxes are from the first qubit and the other two are from the second qubit. For example, if the Box1,4 and Box2,3 are initially even for both qubits, the joint-parity measurement of Box1,2 in the first qubit and Box1,2 in the second qubit gives the following projection:
\begin{multline}
\overbrace{\square\overbrace{\square\square}\square}|\overbrace{\square\overbrace{\square\square}\square}\rightarrow\\
\overbrace{\square\square}\overbrace{\square\square}|\overbrace{\square\square}\overbrace{\square\square}+\overbrace{\Circle\Circle}\overbrace{\Circle\Circle}|\overbrace{\Circle\Circle}\overbrace{\Circle\Circle}.
\end{multline}
where we assume the measurement outcome is even, and the plus sign indicates two classical uncertain states with equal probability. This is an analog to joint-measurement $ZZ$ in the quantum theory.

Meanwhile, for another example with 6 MZMs in the initial state $|0_{12}0_{35}0_{46}\rangle$, we first joint-measure MZM $\gamma_{2}\gamma_{3}\gamma_{4}\gamma_{5}$ and assume even result, we get $(|0_{16}0_{23}0_{45}\rangle+|0_{16}1_{23}1_{45}\rangle)/\sqrt{2}$. Then the measurement of MZM $\gamma_{2}\gamma_{5}$ will reach a definite even result $|0_{16}0_{25}0_{34}\rangle$ due to the cancellation of quantum interference. However, as shown in Fig.\ref{fig:Topology} (d), the HV theory \uppercase\expandafter{\romannumeral2} fails to capture this process and reaches $(\overbrace{\square\overbrace{\square\square}\overbrace{\square\square}\square}+\overbrace{\square\overbrace{\Circle\Circle}\overbrace{\Circle\Circle}\square})/2$ . We can look at another example in Majorana system--the CNOT gate of two MZM qubit which involve both joint-measurement and two-box measurement. We show in the part \uppercase\expandafter{\romannumeral3} of SI\cite{supp} that our HV theory \uppercase\expandafter{\romannumeral2} can not capture CNOT gate process. It is clear that any HV theory can not capture the complete signature of quantum interference. Therefore, we conclude, to reveal certain signatures of quantum interference, the combination of joint-measurement and two box measurement are helpful and necessary.

{\em Discussions--}
There are some other tests for more properties of Majorana systems.
1). Braiding and measurement only cannot generate T-gate, which however can be generated by preparing a noisy magic state ancilla along with state distillation\cite{BravyiKitaev}. T-gate test along with other Clifford operations specifies quantum property which could generate arbitrary quantum states.
2). Bell nonlocality tests clearly complete the Majorana system validation but require a fully universal quantum gate-set including T-gate to break Bell bound. However, the braiding and measurement only cannot break the Bell bound~\cite{dengmajoranagenerator,ClarkePRX,romito2017ubiquitous}. In the future, we need either to fabricate more sophisticated devices along with complicated experimental procedures or to propose novel and simple test schemes.\\

\noindent\textbf{Acknowledgements}

The authors acknowledge the support from NSF-China (Grant No.11974198) and the support from Beijing Academy of Quantum Information Sciences. D.E.L. also acknowledges the support from Thousand- Young-Talent program of China.

\bibliographystyle{apsrev4-1}
\bibliography{toymodel,DissipativeMaj}

\onecolumngrid
\newpage


\section*{Supplementary material}

# Supplementary Information for
## "A hierarchy in Majorana non-abelian tests and hidden variable models"

In this supplementary information, we will provide some details about: I. General fusion procedures in more-Majorana cases. II. General braiding operations in HV theory II. III. Failure in capturing C-NOT gate in HV theory II.

### I. GENERAL FUSION IN MORE MZMS CASE.

First of all, we show the six eigenstates of three Pauli operators in standard quantum case and their corresponding box-states in our HV theories. In standard description of 4-MZM states, we choose $|0_{12}0_{34}\rangle$ as the even eigenstate of both parity measurement $i\gamma_1\gamma_2$ and $i\gamma_3\gamma_4$, where $\gamma_1 = c_{12}^\dagger + c_{12}$, $\gamma_2 = (c_{12}^\dagger - c_{12})/i$, $c_{12}$ is the fermion annihilation operator of the $\gamma_1 - \gamma_2$ pair (similarly for $\gamma_3 - \gamma_4$ pair). Then under basis transformation discussed in the main text, $|0_{14}0_{23}\rangle = (|0_{12}0_{34}\rangle + |1_{12}1_{34}\rangle)/\sqrt{2}$. This is exactly the transformation between different fusion channels, which can be described by $F-$ Matrices. Then we see that $|0_{14}0_{23}\rangle$ is the even eigenstate of parity measurement $X = i\gamma_1\gamma_4$ from the relations mentioned above. Similarly, the even eigenstate of parity measurement $Y = i\gamma_1\gamma_3$ is $|0_{13}0_{24}\rangle = (|0_{12}0_{34}\rangle + i|1_{12}1_{34}\rangle)/\sqrt{2}$. As shown in Fig.S1, we can find that measurement $Z = i\gamma_1\gamma_2$ in Majorana system corresponds to our Box1,2 parity measurement in our HV theories, where this measurement with even parity outcome reaches the state ▢▢ ▢▢. The measurement $X = i\gamma_1\gamma_4$ corresponds to our Box1,4 parity measurement, where the even parity outcome leads to the state ▢ ▢▢ ▢. Similar conclusion also hold for the measurement $Y = i\gamma_1\gamma_3$. In addition, when we apply Box1,2 measurement on ▢ ▢▢ ▢, we get ▢▢ ▢▢ or ◯◯ ◯◯ with equal probability, which correponds to $|0_{14}0_{23}\rangle = (|0_{12}0_{34}\rangle + |1_{12}1_{34}\rangle)/\sqrt{2}$ in quantum case. We can find all parity measurement and basis transformations are similiar with quantum one. We summarized those six eigenstates of three kinds of Pauli matrices (Majorana system) and their corresponding box-pair measurements (HV theories) in Fig.S1.

| (a). States in HV theory | | (b). Corresponding States in Majorana system |
|---|---|---|
| ▢▢ ▢▢ / ◯◯ ◯◯ | $\sigma_z$ | $|0\rangle \rightarrow |0_{12}0_{34}\rangle \;/\; |1\rangle \rightarrow |1_{12}1_{34}\rangle$ |
| ▢▢▢▢ / ◯◯◯◯ | $\sigma_x$ | $|+\rangle = (|0\rangle + |1\rangle)/\sqrt{2} \rightarrow |0_{14}0_{23}\rangle$<br>$|-\rangle = (|0\rangle - |1\rangle)/\sqrt{2} \rightarrow |1_{14}1_{23}\rangle$ |
| ▢▢▢▢ / ◯◯◯◯ | $\sigma_y$ | $|+i\rangle = (|0\rangle + i|1\rangle)/\sqrt{2} \rightarrow |0_{13}0_{24}\rangle$<br>$|-i\rangle = (|0\rangle - i|1\rangle)/\sqrt{2} \rightarrow |1_{13}1_{24}\rangle$ |

Figure S1. The six eigenstates of three Pauli operators in our HV theories and the corresponding explicit form in standard quantum case

In the 4-MZM case, two-box measurements in HV theories are the analog of the fusions in quantum case as shown in the main text. Here, we give more general examples of 6-MZM or more-MZM cases. Considering an example of 6-MZM fusion in quantum case, we measure MZM3,4 on $|0_{12}0_{36}0_{45}\rangle$ and get results $|0_{12}0_{34}0_{56}\rangle$ or $|0_{12}1_{34}1_{56}\rangle$ with equal possibility. Next we measure MZM2,3 on $|0_{12}0_{34}0_{56}\rangle$ and get $|0_{14}0_{23}0_{56}\rangle$ and $|1_{14}1_{23}0_{56}\rangle$ with equal possibility. Or we measure MZM2,3 on $|0_{12}1_{34}1_{56}\rangle$ and get $|0_{14}1_{23}0_{56}\rangle$ and $|1_{14}0_{23}0_{56}\rangle$ with equal possibility. The corresponding procedures in out HV theory is shown in Fig.S2. Suppose the initial state is ▢▢ ▢ ▢▢ ▢, we apply the Box3,4 measurement, and then we can obtain either ▢▢ ▢▢ ▢▢ or ▢▢ ◯◯ ◯◯ with equal possibility, which is the same as those in quantum case. If we obtain the even result ▢▢ ▢▢ ▢▢, the further Box2,3 measurement yields either ▢ ▢▢ ▢ ▢▢ or ◯ ◯◯ ◯ ▢▢. On the other hand, If we obtain the odd result ▢▢ ◯◯ ◯◯ in the first Box3,4 measurement, the further Box2,3 measurement includes a box-pair with a square

and a circle. Here we emphasize that although we know the parity of both Box1,2 and Box3,4 after the previous measurement, the parity of each box is always unknown and is a hidden variable; and therefore, the further Box2,3 measurement leads to either $\square \overbrace{\bigcirc\bigcirc \square \bigcirc\bigcirc}$ or $\bigcirc \overbrace{\square\square \bigcirc \bigcirc\bigcirc}$ with equal probability, which also make sure the total parity of the 4-box is odd. This is also consistent with the measurement outcome from the corresponding quantum case. In more MZMs fusion tests ($2n$ MZMs states), the procedure only affects the fusion channel involved in parity measurement that is at most two pairs (four MZMs), but not affects the other $n$-2 pairs. Therefore we can use the analysis above for the fusion processes in more MZM-cases. We conclude that the fusion tests in 6-MZM and more MZMs in our HV theory show the same measurement consequence as those in the quantum case.

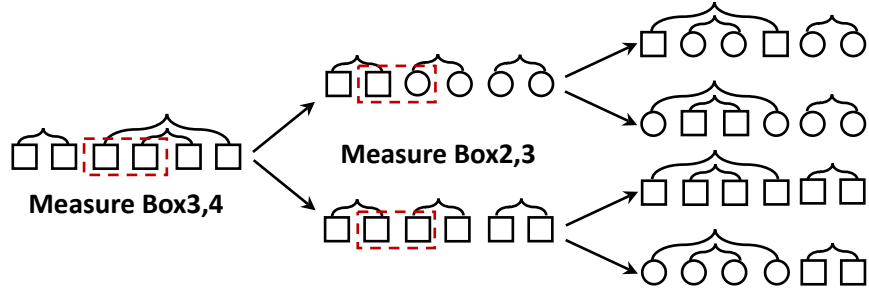

Figure S2. We show an example of fusion in 6-Box case, where we successively measure Box3,4 and Box2,3 and get final result which is consistent with quantum case.

## II. GENERAL BRAIDING OPERATIONS IN HV THEORY II

In this section, we first discuss more examples for the braiding processes applying on various states. Two successive Box2,3 braidings $(P_{23})^2$ on various states are shown in the left part of Fig.S3 (a). The operation $(P_{23})^2$ on the three different box-states corresponds to $\sigma_x$ operation on the $+1$ eigenstate of three Pauli operators in quantum Majorana case, which shows the following transformation:

$$\sigma_x|0\rangle \propto B_{23}^2|0\rangle \propto B_{23}|-i\rangle \propto |1\rangle \tag{S1}$$
$$\sigma_x|+\rangle \propto B_{23}^2|+\rangle \propto B_{23}|+\rangle \propto |+\rangle \tag{S2}$$
$$\sigma_x|+i\rangle \propto B_{23}^2|+i\rangle \propto B_{23}|0\rangle \propto |-i\rangle \tag{S3}$$

We can also examine two successive Box1,2 braidings on various states as shown in the right part of Fig.S3 (a). These correspond to $\sigma_z$ operation on the $+1$ eigenstate of three Pauli operators, which shows the following transformation:

$$\sigma_z|0\rangle \propto B_{12}^2|0\rangle \propto B_{12}|0\rangle \propto |0\rangle \tag{S4}$$
$$\sigma_z|+\rangle \propto B_{12}^2|+\rangle \propto B_{12}|+i\rangle \propto |-\rangle \tag{S5}$$
$$\sigma_z|+i\rangle \propto B_{12}^2|+i\rangle \propto B_{12}|-\rangle \propto |-i\rangle \tag{S6}$$

With the discussion of the single braiding and the two successive braiding above, we can generalize the discussion to the general multiple braidings. Therefore, we see that braiding operations in our HV theory II generate the same measurement outcomes as those in the Majorana quantum systems. Another example is Hadamard gate $P_{12}P_{23}P_{12}$ on various states as shown in Fig.S3 (b). We see that Hadamard gate in our HV theory II corresponds to those in Majorana systems:

$$H|0\rangle = |+\rangle, H|+\rangle = |0\rangle, H|i\rangle \propto |-i\rangle, \tag{S7}$$

**(a). Successive Box2,3 and Box1,2 braidings on various states**

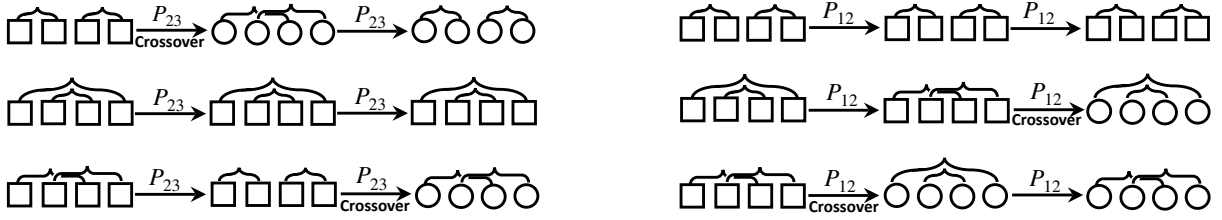

**(b). Hadamard gate on various states**

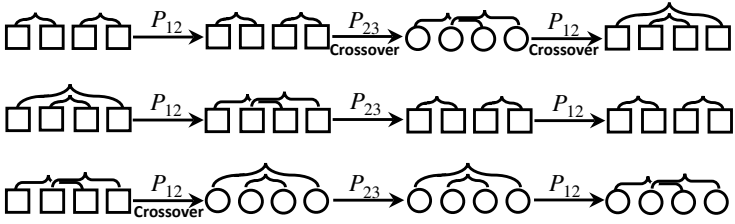

**(c). Braiding in 6-MZM**

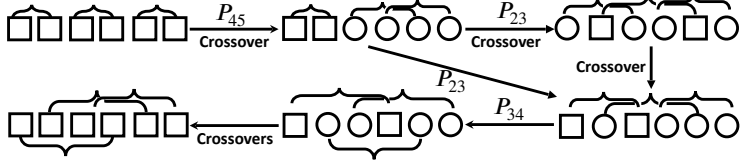

**(d). CNOT gate in dense code(6-MZM)**

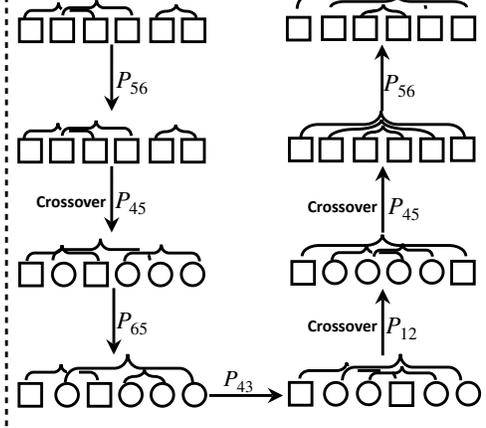

Figure S3. We complete discussion of braiding and Hadamard gate in HV theory II, where we successively braid Box2,3 and Box1,2 on various states and get final result which is consistent with quantum case. We also show an example of braiding in dense coding with 6-MZM and CNOT gate in 6-MZM case only with braiding operation

We further discuss the braidings in cases with more than 4 MZMs. In either Majorana model or our HV theory II, the objects (MZMs or boxes) describing a specific fusion channel are all paired. Therefore, the braiding of two objects involves at most two pairs that is four MZMs or boxes. Let's now focus on our HV theory II. Considering the braiding of two boxes among $2n$ boxes, this procedure only changes the topology of two connections (of two box-pairs) involved in the braiding, but can not change the relative positions between these two connections and the other $n-2$ connections. For example in Fig.S3(c), for initial state ▭▭ ▭▭ ▭▭, we first apply the anticlockwise braiding of Box4,5 and then a crossover; we get ▭▭ ○ ○○○. Then, we apply the anticlockwise Box2,3 braiding and then a crossover, and we get ○▭○ ○▭○. The corresponding quantum case $B_{23}B_{45}|0_{12}0_{34}0_{56}\rangle \propto |1_{13}0_{25}1_{46}\rangle$ gives the same result as our case. Finally, we want to apply the Box3,4 braiding in this example. Considering the analysis of this step, we first apply a crossover to move the Box1,3's connection in front of the Box2,5's connection; then we braid Box3,4 and finally convert the state to the standard form. The final result of this example has the same measurement consequence as those in the quantum case $B_{34}B_{23}B_{45}|0_{12}0_{34}0_{56}\rangle \propto |0_{14}0_{25}0_{36}\rangle$. Similar to this example, any braiding step in more box-cases involves only four boxes, therefore we can use the analysis above for braiding procedure in more box-cases. We conclude that the braiding procedures in multiple-box setup show the same measurement consequence as those in the Majorana models.

Finally, we show one more example–the CNOT gate in dense coding with 6-MZM case[1] which only requires braiding procedures. In dense coding of six MZMs, we define the logical two-qubit basis in fixed total even parity as:

$$|00\rangle_L \equiv |0_{12}0_{34}0_{56}\rangle, |01\rangle_L \equiv |0_{12}1_{34}1_{56}\rangle, |10\rangle_L \equiv |1_{12}1_{34}0_{56}\rangle, |11\rangle_L \equiv |1_{12}0_{34}1_{56}\rangle. \tag{S8}$$

We treat the first qubit as the controlled qubit and the second qubit as the target qubit. Applying a sequence of Majorana braiding operations, we have

$$B_{56}B_{45}B_{12}B_{43}B_{65}B_{45}B_{56}|0_{13}0_{24}0_{56}\rangle \propto |0_{15}0_{26}0_{34}\rangle. \tag{S9}$$

which can be converted into the logical two-qubit basis:

$$B_{56}B_{45}B_{12}B_{43}B_{65}B_{45}B_{56}(|00\rangle_L + i|10\rangle_L)/\sqrt{2} \propto (|00\rangle_L + i|11\rangle_L)/\sqrt{2}, \tag{S10}$$

which creates the entanglement. Obviously operation $B_{56}B_{45}B_{12}B_{43}B_{65}B_{45}B_{56}$ is equivalent to CNOT in this basis definition. In our HV theory II we begin with initial state: ⊡⊡⊡⊡ ⊡⊡, and then successively apply the braiding operations: $P_{56}, P_{45}, P_{65}, P_{43}, P_{12}, P_{45}, P_{56}$ to get ⊡⊡ ⊡⊡ ⊡⊡, which is consistent with quantum case, as shown in Fig.S3 (d).

### III. FAILURE ON C-NOT GATE IN HV THEORY

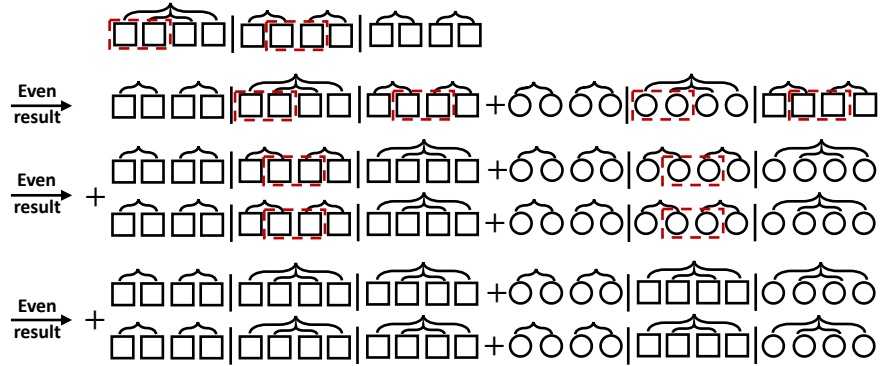

Figure S4. (a) Conventional CNOT gate where each four MZMs form a single physical qubit. In this case, the extra ancillary qubit along with joint-measurements are necessary to generate CNOT. (b) The failure to capture this conventional CNOT gate in our HV theory.

As the discussion in the previous section, the CNOT in dense coding with 6 MZMs contains only braiding operations; and therefore, this gate can be fully captured and show exactly the same results as those in our HV theory. However, in the conventional CNOT, where each four MZMs forms a single physical qubit, requires ancillary qubit and joint-measurement in Majorana systems as shown in Fig.S4 (a) [2]. As shown in the main text, along with the joint-measurement, the fusion and braiding procedures in the HV theory cannot reproduce the quantum interference behaviors in real Majorana quantum systems. Here, we show explicitly how our HV theory fail to capture the conventional CNOT operations as shown in Fig.S4 (b). We begin with initial state:

$$\square \overbrace{\square\square} \square | \overbrace{\square\square} \overbrace{\square\square} | \overbrace{\square\square} \overbrace{\square\square}, \tag{S11}$$

Apply joint-measurement of $ZX$ on 1st and 2ed bit and assume even result, then we get:

$$\overbrace{\square\square} \overbrace{\square\square} | \square \overbrace{\square\square} \square | \overbrace{\square\square} \overbrace{\square\square} + \overbrace{\circ\circ} \overbrace{\circ\circ} | \circ \overbrace{\circ\circ} \circ | \overbrace{\square\square} \overbrace{\square\square}, \tag{S12}$$

with equal probability for each outcome. Then we measure $ZX$ on the second and the third "qubit" and still assume even result, then we get:

$$\overbrace{\square\square} \overbrace{\square\square} | \overbrace{\square\square} \overbrace{\square\square} | \square \overbrace{\square\square} \square + \overbrace{\square\square} \overbrace{\square\square} | \overbrace{\circ\circ} \overbrace{\circ\circ} | \circ \overbrace{\circ\circ} \circ \tag{S13}$$

$$+ \overbrace{\circ\circ} \overbrace{\circ\circ} | \overbrace{\square\square} \overbrace{\square\square} | \square \overbrace{\square\square} \square + \overbrace{\circ\circ} \overbrace{\circ\circ} | \overbrace{\circ\circ} \overbrace{\circ\circ} | \circ \overbrace{\circ\circ} \circ, \tag{S14}$$

where the probability for each outcome is $1/4$. Here, however in quantum case, we actually get a negative sign in front of the 4th term, which will generate totally different result later. Switching back to this HV model, if we continue to measure $X$ in the 2nd "qubit" and assume even result, we get

$$\overbrace{\Box\Box}\;\overbrace{\Box\Box}\;|\Box\overbrace{\Box\Box}\Box\rangle|\Box\overbrace{\Box\Box}\Box\rangle + \overbrace{\Box\Box}\;\overbrace{\Box\Box}\;|\Box\overbrace{\Box\Box}\Box\rangle|\circ\overbrace{\circ\circ}\circ\rangle \qquad (S15)$$

$$+ \overbrace{\circ\circ}\;\overbrace{\circ\circ}\;|\Box\overbrace{\Box\Box}\Box\rangle|\Box\overbrace{\Box\Box}\Box\rangle + \overbrace{\circ\circ}\;\overbrace{\circ\circ}\;|\Box\overbrace{\Box\Box}\Box\rangle|\circ\overbrace{\circ\circ}\circ\rangle, \qquad (S16)$$

which is the finial result of CNOT gate in the HV theory. However, in quantum case, the quantum interference will cancel terms and generate simply $|0\rangle|+\rangle|0\rangle$ corresponding to the first two terms here in out HV model. Therefore, the quantum interference will lead to the final state is $(|0\rangle|+\rangle|0\rangle + |1\rangle|+\rangle|1\rangle)/\sqrt{2}$, where the first and the third qubit are entangled. However, in our HV theory, the measurement of the first and the third "qubits" will not get correlated results. Therefore, HV theory cannot reproduce the results from quantum interference in Majorana systems.

At last, we give a table which summarize the conclusion we get in the main text.

| Test \ Property | Non-locality | Topology | Quantum Interference |
|---|---|---|---|
| **Only Fusion** | ✓ | ✗ | ✗ |
| **With braiding** ★ Hadamard gate ★ Successive braiding | ✓ | ✓ | ✗ |
| **With Joint-Measurement in more MZMs** | ✓ | ✓ | ✓ |

Figure S5. We conclude our results in this table. For each test we give a name and we use tick or cross to show our conclusion on whether this test can show distinct property from comparison of our models and standard one.



\end{document}